\documentclass[preprint,11pt,authoryear]{elsarticle}

\usepackage{amssymb}
\usepackage{amsmath}
\usepackage{braket}

\usepackage[margin=1in]{geometry}

\newcommand \be{\begin{equation}}
\newcommand \ee{\end{equation}}

\newcommand \bea{\begin{eqnarray}}
\newcommand \eea{\end{eqnarray}}

\newcommand \bse{\begin{subequations}}
\newcommand \ese{\end{subequations}}

\newcommand \bml{\begin{subequations}\begin{eqnarray}}
\newcommand \eml{\end{eqnarray}\end{subequations}}

\newcommand \nn{\nonumber}

\newcommand \mcE{{\mathcal E}}

\begin{document}

\begin{frontmatter}



\title{Neutral atom quantum computing} %% Article title


\author{Mark Saffman} %% Author name

\affiliation{organization={Department of Physics, University of Wisconsin-Madison, Madison, WI 53706, USA \\ Infleqtion, Madison, WI, 53703, USA}
%Department and Organization
 %           addressline={} 
%            city={Madison},
        %    postcode={53706}, 
 %           state={Wisconsin},
  %          country={USA}
  }

%% Abstract
\begin{abstract}
Neutral atom qubits are one of the leading approaches for implementation of a large scale quantum computer. The original proposals for neutral atom qubits were formulated more than 25 years ago, with first demonstrations of a universal gate set following 10 years later. In the last few years the performance and scale of neutral atom qubit arrays has developed at a rapid pace leading to demonstrations of quantum algorithms, and logical qubits for fault tolerant error correction. This contribution reviews the physics of the neutral atom approach, surveys current capabilities, and provides an outlook for future progress.
\end{abstract}



%% Keywords
\begin{keyword}


neutral atom qubits
\sep 
quantum computing
\sep 
optical traps
\sep
Rydberg atoms
\sep
quantum error correction
\end{keyword}

\end{frontmatter}



\tableofcontents

%\listoffigures

%\listoftables

\section{Introduction}
\label{sec1}

Quantum computers are information processing devices that achieve enhanced computational power compared to conventional machines by leveraging quantum mechanical principles of nature (\cite{Feynman1982,Feynman1986, Manin1980, Deutsch1985}). In particular, quantum superposition enables objects with two discrete levels (qubits) to be in a superposition state $\ket{\psi}=c_0\ket{0}+c_1\ket{1}$ where $c_0, c_1 $ are arbitrary complex coefficients, subject only to normalization $|c_0|^2+|c_1|^2=1$. A collection of $N$ qubits can represent a superposition of $2^N$ different quantum states, thereby providing the potential for massively parallel information processing.  Accessing the power of a quantum computer is not  simple, as a measurement of the output of a computation collapses the quantum superposition to a single state out of $2^N$ possibilities. Nevertheless, carefully designed quantum algorithms exploit interference between different states to enhance the probability of observing an output that encodes a useful computational result (\cite{Nielsen2000}). 

The other crucial ingredient in a quantum computer is entanglement. While classical wave interference can be used to {\it simulate} a quantum computer (\cite{Cerf1998, Spreeuw2001}) careful analysis shows that without entanglement, resource requirements in energy, space, or time always grow exponentially as a function of the problem size for any classical simulation (\cite{Ekert1998}). Thus a useful  quantum computer requires qubits that can be prepared in superposition states which are protected from decoherence throughout the computation. It must also be possible to entangle arbitrary pairs of qubits and, together with control of the state of individual qubits, perform arbitrary unitary transformations on the collective state of all $N$ qubits. Also a method for measuring the quantum state at the end of the calculation must be available. Since small quantum systems with a limited number of qubits can be readily simulated on a conventional computer, systems with a large number of qubits are a requirement for the quantum computer to provide new capabilities.   Taken together there are a set of desiderata  (\cite{DiVincenzo2000})
that must be satisfied by any viable physical implementation of useful quantum computation. 

Satisfying these requirements is extremely challenging and has been a scientific and technological goal for more than thirty years. There are a multitude of physical approaches that are potentially suitable and five of them are being vigorously developed: trapped ions, neutral atoms, superconducting circuits, semiconductor quantum dots, and 
photonics (\cite{Ladd2010,Bergou2021}).
These five approaches can be grouped into two categories. Naturally occurring qubits including trapped ions, neutral atoms, molecules, and also photons. Such natural qubits are identical and also exhibit very long coherence times. In contrast, superconducting circuits and semiconductor quantum dots must be fabricated, and are subject to variations in material properties  when integrating many qubits into a single system. Every qubit modality has a unique set of characteristics, some of which 
are enabling for quantum computation, and some of which present difficulties that must be overcome to realize a large scale quantum processor. In the remainder of this chapter we present the physics underlying our ability to encode qubits in neutral atoms and use them for error corrected, and fault tolerant quantum computation. Given space limitations it is not possible to comprehensively cover all relevant topics and we refer to several review articles for further details 
(\cite{Ryabtsev2005,Saffman2010,Saffman2016,Kaufman2021,Browaeys2020,Morgado2021,XWu2021,XFShi2022}). Also the important topic of quantum simulation with neutral atoms will not be covered here. Comprehensive reviews are available in the literature (\cite{Bloch2012,Altman2021}).

\section{Choice of atomic species}

In order to use atoms as qubits it is most convenient if the atoms can be trapped in well defined locations. Although not an absolute requirement, and groundbreaking quantum experiments have been performed with moving atoms that rapidly pass through an apparatus (\cite{Haroche}), there are many practical advantages to working with stationary atoms. This requires cooling the atoms to a low temperature and low velocity state and trapping them in a way that is compatible with preserving coherence of encoded quantum states.

\begin{figure}[!t]
\center
\includegraphics[width=.8\columnwidth]{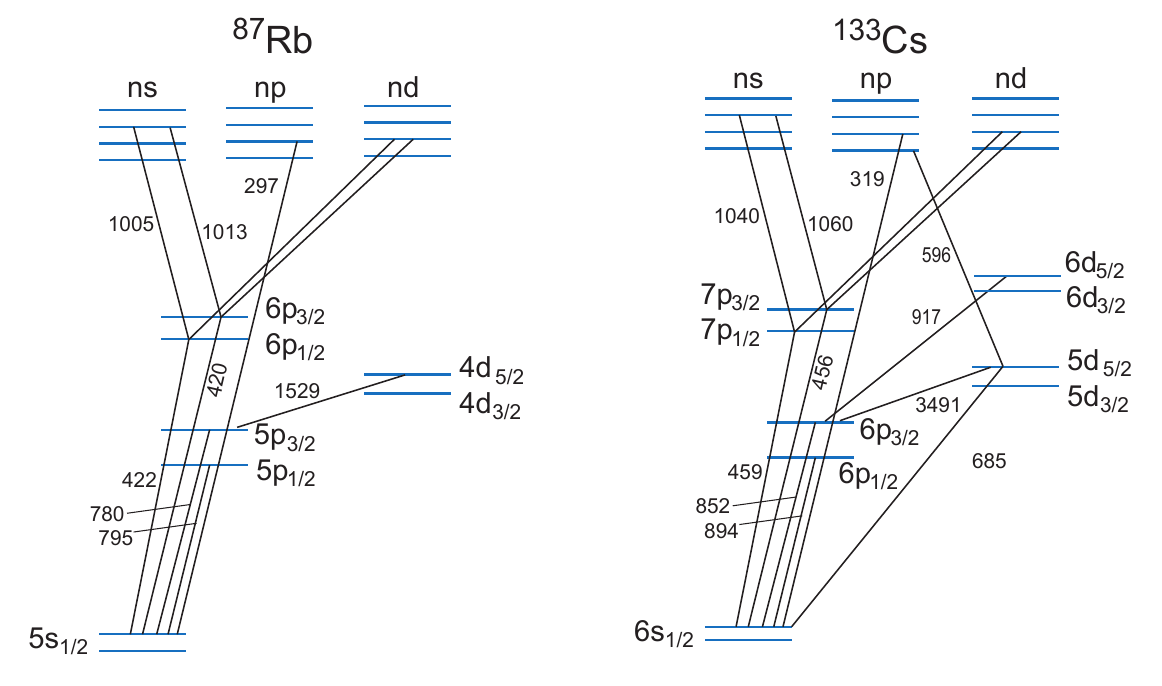}
\caption{\label{fig.alkali_atoms} Level structure and transitions that are of primary interest for quantum computing with Rb and Cs atoms. The Rydberg levels labeled $ns, np,nd$ may have $n$ ranging from about 40-100. Transition wavelengths are shown in nm. 
For Rb the wavelength dependent functions are:  cooling (780), optical pumping (795), two-photon Rydberg excitation (422/1005 \cite{Evered2023}) or (420/1013 \cite{Viteau2011}), one-photon Rydberg (297  \cite{Thoumany2009}), excited state hiding (1529 \cite{BHu2025}). 
For Cs: cooling (852 or 685/3491 \cite{Scott2025}), optical pumping (894), two-photon Rydberg excitation (459/1040 \cite{Graham2019}) or (456/1060 \cite{Anand2024}) or (685/596 \cite{Bohorquez2023}), one-photon Rydberg (319 \cite{Jau2016}), excited state hiding (917). In addition a range of wavelengths are used for trapping and atom transport.  }
\end{figure}

These requirements can be readily achieved by combining laser cooling with optical trapping techniques. Laser cooling has been demonstrated on many different atomic elements.  The species that have reached an advanced level of development for quantum computing are the alkali atoms Rb and Cs, the alkaline earth Sr and the alkaline earth like lanthanide Yb. We will henceforth  refer to both Sr and Yb as alkaline earth like (AEL) atoms. 
The level structure and relevant wavelengths used for quantum computing with the heavy alkali atoms Rb and Cs are shown in Fig. \ref{fig.alkali_atoms}. Also Li atoms have been used experimentally (\cite{Guardado-Sanchez2018}).   The wavelengths span a wide range from the ultraviolet near 300 nm for one-photon excitation, to the infrared near 1000 nm for two-photon  excitation, and near 1500 nm for hiding operations. Not shown in the figure are non-resonant wavelengths used for optical trapping. These may vary from the visible to the infrared depending on the atom and the type of trap.

\begin{figure}[t!]
\centering\includegraphics[width=.8\textwidth]{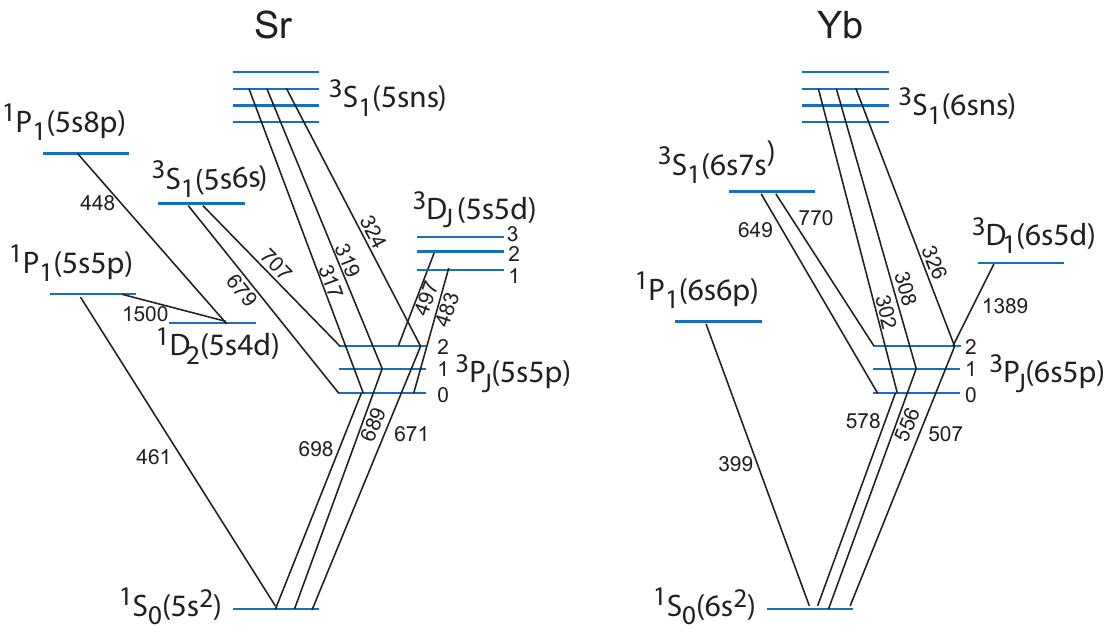}
\caption{Level diagrams and often used wavelengths in nm for Sr and Yb  atoms. For Sr the wavelength dependent functions are:  
cooling (461, 689), clock transitions (671, 698), repumpers (448, 679, 707, 483, 497), Rydberg excitation (317, 319, 324). 
For Yb: 
cooling (399,556), clock transitions (507,578), repumpers (649,770),
Raman transitions (649,770), Rydberg excitation (302, 308, 326). 
 In addition a range of wavelengths are used for trapping and atom transport. }
\label{fig.alkaline_wavelengths}
\end{figure}

Wavelengths used for qubits encoded in the AEL atoms Sr and Yb are shown in Fig. \ref{fig.alkaline_wavelengths}. 
As is the case for the alkali atoms there is a wide range of wavelengths stretching from near 300 nm to about 1500 nm. 
In addition, the lanthanide atoms Er, Dy, Ho, Tm are starting to be used for quantum information experiments (\cite{Hostetter2015,Golovizin2015,Trautmann2021,Bloch2023}), presenting yet another set of wavelengths. 

\section{Atom cooling and trapping}

Laser cooling techniques are widely used in experimental atomic physics (\cite{Footbook,vanderStratenbook}). There are a variety of methods that can be brought to bear ranging from relatively simple Doppler cooling to Raman methods that require more sophisticated optical control but can reach lower temperatures. Using Doppler cooling techniques atoms  can be rapidly cooled  to temperatures of $10-1000~\mu\rm K$ depending on the atomic species. Further cooling using polarization gradient methods, or Raman cooling can lower the temperature to the few $\mu\rm K$ range or below. Standard thermometry is not useful for characterizing such low temperatures. Instead temperature is defined in terms of the velocity distribution as 
\be 
k_{\rm B}T=\frac{m\langle v^2\rangle}{3}
\ee
where $k_{\rm B}$ is the Boltzmann constant, $m$ is the atomic mass, and $v^2$ is the square of the three dimensional velocity. A Cs atom at $T=1~\mu\rm K$ has root mean square speed  $\sqrt{\langle v^2\rangle }=1.4~\rm cm/s. $

The most developed method of trapping pre-cooled atoms is to use far-detuned optical traps. Provided the trap depth is large compared to the translational energy $k_{\rm B}T$ the atom can be effectively confined.  The general situation is that a laser beam connects ground states $\ket{g_0}, \ket{g_1}$ to an excited state $\ket{e}$ with a large detuning $\Delta=\omega-\omega_a$ where $\omega$ is the frequency of the light and  $U=\hbar\omega_a$ is the transition energy , with $\hbar$ Planck's constant. Population in the excited state decays at a rate $\Gamma$. The coupling rate between the ground and excited atomic states is given by the Rabi frequency $\Omega=d_{\rm eg} {\mathcal E}/\hbar$, $\mathcal E$ is the electric field amplitude, and $d_{\rm eg}=\bra{e}\hat d\ket{g}$ is the transition matrix element of the atomic dipole operator $\hat  d$.  
 For simplicity we assume $d_{\rm eg_{0}}=d_{\rm eg_{1}}\equiv d_{\rm eg}.$ In the general case the energy shift of atomic ground states is found in second order perturbation theory by summing the coupling to all excited states. Doing so results in a frequency dependent  polarizability $\alpha(\omega)$ and an energy shift of the ground state given by 
\be
U(\omega)=-\frac{1}{4}\alpha(\omega) |\mcE|^2.
\label{eq.alpha}
\ee
Equation (\ref{eq.alpha}) encompasses only the simplest case of a scalar polarizability. More generally there can be vector and tensor contributions which result in the energy shift having a dependence on the polarizability of the light (\cite{Mitroy2010,LeKien2013}).

When $\Delta\ll \omega, \omega_a$ we can make a rotating wave approximation and derive simple analytical expressions for the essential features of the light-atom interaction. 
This idealized two-level model of an atom interacting with coherent light is treated in many books (see e.g. \cite{Mandel1995}).
The effect of the laser light on the atom manifests itself via coherent and incoherent processes. The coherent process is a shift of the energy of the ground states by amounts  
\be 
\delta U_0 =  \hbar\frac{|\Omega|^2}{4(\Delta-\omega_{\rm q}/2)},~~~~ \delta U_1 =\hbar  \frac{|\Omega|^2}{4(\Delta+\omega_{\rm q}/2)}. 
\ee 
When $\Delta\gg \omega_{\rm q}$ we find average and  differential light shifts of 
\be 
U_{\rm g} =\frac{\delta U_0 +\delta U_1 }{2} =\hbar \frac{|\Omega|^2}{4\Delta},~~~~ U_{10} =  \delta U_1 -\delta U_0 =-\hbar\frac{|\Omega|^2 \omega_{\rm q}}{4\Delta^2}. 
\label{eq.DLS}
\ee 
The average energy shift $U_{\rm g}$ is positive for $\Delta>0$ which is referred to as blue detuning, and is negative for $\Delta<0$ which is called red detuning. For large detunings with $|\Delta|> \omega_{\rm q}/2$ the differential shift $U_{10}$ is always negative, independent of the sign of $\Delta.$ The ratio of the differential shift to the average shift is $\left|U_{10}/U_{\rm g}\right|=\omega_{\rm q}/\Delta.$ For $^{87}$Rb and $^{133}$Cs the qubit frequencies are 6.8 and 9.2 GHz, while the detunings used for atom trapping are typically 10-100 THz. Thus the differential shift is a small  fraction of the average shift and far detuned light effectively traps both qubit states. Nevertheless, for the longest possible coherence times it is of interest to minimize the DLS which can be done with so-called ``magic'' trapping techniques (see Sec. \ref{sec.magic}).

The incoherent process is scattering of photons from the excited state. The scattering rate at short times and large detuning  is approximately 
$ 
r=\frac{\Gamma^3}{4\Delta^2}\frac{I}{I_{\rm sat}}
$ 
where the optical intensity is $I=\frac{\epsilon_0 c}{2}|\mcE|^2$ and $I_{\rm sat}$ is the two-level saturation intensity. Using  $I_{\rm sat}=2\pi^2\hbar c\Gamma/(3\lambda_{eg}^3)$ with $\lambda_{eg}$ the transition wavelength we can express the scattering rate as \footnote{This expression for the scattering rate is valid at short times for which the probability of having scattered a photon is much smaller than unity. At longer times the average scattering rate is $1/2$ this amount. See \cite{Grimm2000} for a more detailed derivation.}
\be
r=\frac{3\lambda_{eg}^3}{8\pi^2 \hbar c}\frac{\Gamma^2}{\Delta^2} I.
\ee
 We see that the scattering rate decreases as $1/\Delta^2$ whereas the light shift scales as $1/|\Delta|$. Thus at large detunings we obtain an effectively coherent light-atom interaction with minimal scattering.

\begin{figure}[t!]
\centering\includegraphics[width=.7\textwidth]{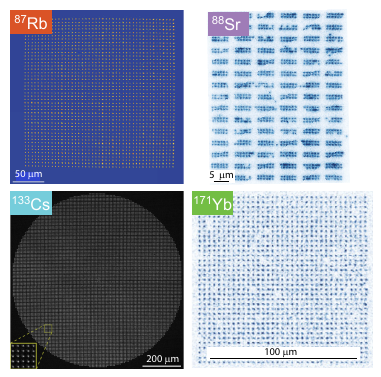}
\caption{Demonstrations of large qubit arrays. $^{87}$Rb: 
Nearly defect free
square array with 2024 atoms in 2025 sites (\cite{RLin2025});
$^{133}$Cs: averaged image of atoms in a 12,001 site array  (\cite{Manetsch2025});
$^{88}$Sr: more than 1000 atoms loaded into an optical lattice (\cite{Gyger2024});
$^{171}$Yb: 1225 site array with 99\% filling fraction loaded from a reservoir   (\cite{Norcia2024}).
}
\label{fig.arrays}
\end{figure}

\subsection{Bright traps}

To make a bright trap an atom in state $\ket{g}$ is illuminated by a focused lowest order Gaussian beam with intensity 
\begin{equation}
    I(r,z)= \frac{2P}{\pi w^2(z)}e^{-2 r^2/w^2(z)}.
\end{equation}
Here we assume a radially symmetric beam, $z$ is the propagation coordinate, $r$ is the transverse distance from the beam axis, $w^2(z)=w_0^2(1+z^2/L_R^2)$ where $w_0$ is the beam waist ($1/e^2$ intensity radius) located at $z=0$ and $L_{\rm R}= \pi w_0^2/\lambda$ with $\lambda$ the optical wavelength. The total power in the beam is $P$. 
Provided the detuning is negative, or more generally the dynamic polarizability $\alpha(\omega)$ is positive,   the light shift in a  focused Gaussian beam provides an attractive three-dimensional potential. In the more general case of large detunings where the rotating wave approximation is no longer valid the requirement for an attractive potential is that the dynamic polarizability $\alpha(\omega)$ is positive.

An atom trapped in this potential will be localized in space around the origin $r=z=0$ with variances (\cite{Saffman2005a}) 
\begin{eqnarray}
\langle r^2\rangle &=& \frac{w_0^2}{4}\frac{k_{\rm B} T}{U_0},\\
\langle z^2\rangle &=& \frac{\pi^2w_0^4}{2\lambda^2}\frac{k_{\rm B} T}{U_0}.
\end{eqnarray}
Here  $T$ is the atom temperature and $U_0$ is the light shift at trap center.  Temperature to trap depth ratios of $k_{\rm B} T/U_0 < 0.01$ are routinely achieved with $w_0\sim \lambda$ so the position variance in all three dimensions  can be as small as a few percent of the trap waist, which corresponds to a few tens of nanometers. Large arrays of tweezers with more than 10,000 trap sites have been created using spatial light modulators (\cite{Manetsch2025}), and with more than 100,000 sites using metasurface optical components (\cite{Holman2026}). Representative examples for the most used atomic elements are shown in Fig. \ref{fig.arrays}.

\subsection{Dark traps}
\label{sec.darktraps}

It is also possible to create so-called ``bottle'' traps that localize atoms at a minimum of the optical intensity. Choosing the wavelength of the trapping light such that the polarizability is negative, the light induced potential becomes repulsive which traps atoms in a dark spot surrounded by light.
Using bottle traps has some advantages in that the atomic qubits are exposed to a much lower intensity of trap light. This tends to increase the coherence time and reduces level shift effects that can complicate laser cooling and  quantum state control operations.

There are many different ways in which bottle traps, and arrays of bottle traps can be created, including designs with active or passive components (\cite{Xu2010,Li2012,Huft2022}).  
Nevertheless, bottle type traps have not been as widely adopted as attractive tweezer traps. There are two main reasons for this. The optical systems required to form arrays of bottle traps tend to be more complex than for tweezer arrays. In addition,  bottle traps that are produced using a single structured beam have saddle points in the confining potential that are about $1/3$  as deep as the confinement in the radial and axial directions (\cite{YXiao2021}). This implies that more optical power per trap is required for bottle beams which is undesirable in the context of scaling to very large arrays. 
% {(add bottle figure ??)}

\subsection{Optical lattices}

Optical lattices formed by interference of a few laser beams are an alternative to arrays of tweezer or bottle traps. Lattices can be configured to provide one- two- or three-dimensional trap arrays and have been extensively used for many-body physics studies and quantum simulation \cite{Bloch2012}. Standard optical lattices formed by counter-propagating beam pairs with wavelength $\lambda$ provide traps spaced by $\lambda/2.$  For typical micron scale wavelengths such traps are so closely spaced that  control and measurement of individual atoms is very challenging. 

Approaches based on small angle interference between 
co-propagating beam pairs in 3D (\cite{Nelson2007})
or an arrangement of folded, elliptical 
beams in 2D (\cite{RTao2024}) have been used to provide more convenient site spacings of a few microns. Since each light beam contributes to more than one trap the scaling of the number of trap sites with optical power exceeds the linear scaling of tweezer arrays. Although not yet as well developed as tweezer or bottle arrays, the optical lattice approach may ultimately be a viable alternative for large arrays with greater than 10,000 sites.

\subsection{Atom lifetime}

Optical traps for atoms provide trap potentials with a typical depth measured in temperature units of 
$U/k_{\rm B}\lesssim 1~\rm  mK.$ Although deeper traps are possible, the increase in power requirements and increased scattering rates for tweezer traps make large arrays with trap depths more than a few mK impractical. 
Untrapped atoms and molecules in a vacuum system have kinetic temperatures from collisions with the vacuum system walls that are 300 K for 
room temperature systems, and a few K in a  cryogenic apparatus. Collisions with untrapped particles can eject atoms from the traps leading to a finite qubit lifetime even in cryogenic systems with exceptional vacuum quality. Qubit lifetimes as long as 6000 s have been recorded in a cryogenic apparatus where cryopumping contributed to very low background pressure (\cite{Schymik2021}). To achieve such a long lifetime the atoms were periodically laser cooled  in order to mitigate  heating from glancing collisions and trap intensity noise (\cite{Gehm1998,Bali1999}).

\section{Qubit encoding, intialization, and measurement}

\begin{figure}[t!]
\centering\includegraphics[width=.7\textwidth]{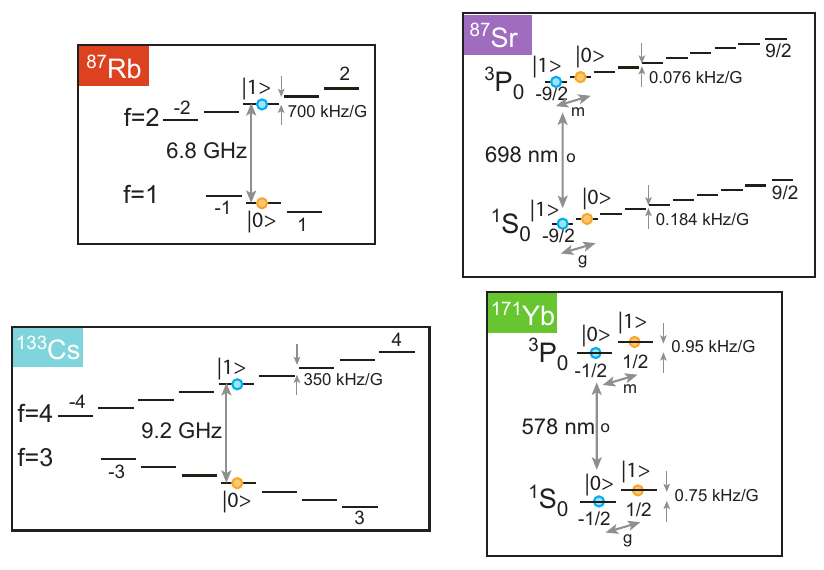}
\caption{Qubit encodings in alkali and AEL atoms. 
The qubit transitions are indicated with gray arrows with the omg choices labeled for Sr and Yb. Rb and Cs use $m_f=0$ clock states that have no linear sensitivity to magnetic fields and a weak quadratic sensitivity. In Sr and Yb there are several {\it omg} choices for encoding. We may use nuclear spin states in the electronic ground or metastable levels with $M_F$ a half integer, and  which have small linear Zeeman shifts. Alternatively an optical transition between ground and metstable levels can be used. 
}
\label{fig.encoding}
\end{figure}

\subsection{Qubit encoding}

Qubits can be encoded in the internal states of single atoms as shown in Fig. \ref{fig.encoding} for alkali and AEL atoms. For the alkali atoms the most widely used encoding is the $m_f=0$ hyperfine-Zeeman clock states with total angular momentum $f=I\pm 1/2$ where $I$ is the nuclear spin. The coherence time of these states can reach many seconds limited only by scattering of the trap light, magnetic field noise, and differential light shifts as described by Eq. (\ref{eq.DLS}). 

AEL atoms present a richer set of choices for encoding in the so-called optical-metastable-ground ({\it omg}) architecture (\cite{NCHen2022,Lis2023}). The fermionic isotopes of AEL atoms have half integer nuclear spin $F$ and spin singlet  electronic ground states $^1{\rm S_0}$  with no orbital or electronic angular momentum. Thus the ground state structure is a single level with total angular momentum $F$ and nuclear Zeeman states with spin projection M satisfying $-F\le M\le F.$ It was first proposed in \cite{Daley2008} to use the nuclear spin register for encoding quantum states. A basic qubit encoding utilizes two of the nuclear spin states in the $^1{\rm S_0}$  ground state. Because the nuclear spin couples only very weakly to external magnetic fields this encoding provides very good coherence properties as we discuss in the next section. 

An alternative encoding uses nuclear Zeeman states in the $^3{\rm P_0}$ metastable state.  In $^{87}$Sr and $^{171}$Yb the ${^3}{\rm P_0}$
states have lifetimes of 167 and 21 s respectively, suitable for long coherence qubit encoding. The optical qubit encoding uses one  
$^1{\rm S_0}$ ground state and one $^3{\rm P_0}$ excited state thereby providing an optical frequency qubit on the ${^1}{\rm S_0} - {^3}{\rm P_0}$  transition which forms the basis of the world's best optical clocks in AEL atoms (\cite{Ludlow2015}).  
Unlike the situation with alkali atom microwave frequency clock qubits, here the qubit frequency is comparable to the detuning of the trapping light. The implication is that the DLS on the qubit states is significant and achieving good coherence and reliable state control requires suppressing or eliminating the DLS. This is done using specific
wavelengths for the trap light such that the polarizability of both qubit states is the same. These so-called ``magic'' wavelengths are explained in more detail in Sec. {\ref{sec.magic}.  In addition to quantum computing applications the combination of single atom tweezer traps with optical clock atoms has led to a new generation of optical tweezer clocks (\cite{Madjarov2019,Norcia2019}).

Finally, a fine-structure THz frequency qubit encoding is also possible using a $^3{\rm P_0}$ state and a metastable $^3{\rm P_2}$ state (\cite{Pucher2024}). This encoding has primarily been explored in bosonic $^{88}$Sr. The use of a bosonic isotope with zero nuclear spin simplifies the level structure so the qubit is now encoded in $M_J=0$ states of the fine structure levels. 

The standard encoding mechanism assigns each qubit to a single atom. It is also possible to use atoms with a large number of stable internal states to encode more than one qubit. In this case control of specific encoded qubits relies on the ability to address internal states either spectroscopically or by selection rules. An example is provided by lanthanide atoms that have large manifolds of hyperfine-Zeeman states in the electronic ground state. The element with the largest number of electronic ground states is Ho which has nuclear spin $I=7/2$ and electronic angular momentum $J=15/2$ in the ground state. This gives a total of 128 Zeeman states (\cite{Saffman2008}). One could imagine encoding a qudit with $d=128$ in the 128 dimensional Hilbert space. This is not useful in any simple way since logical operations would be exceedingly complex. Nevertheless, this type of approach has been demonstrated for a more modest value of $d=13$  using electronically excited, metastable states in trapped ions (\cite{PJLow2025}). The  {\it omg} framework that is used in AEL atoms, see Fig. \ref{fig.encoding}, was first introduced in trapped ions (\cite{Allcock2021}) and analogous qudit encodings could also be considered for metastable states in AEL atoms.

With 128 states available one might naively consider encoding 64 qubits. This is not possible since the Hilbert space spanned by 64 qubits would require an exponentially large atomic state space with  $2^{64}$ states. Although 7 qubits could in principle be encoded in 128 states, implementing a complete set of logical operations would be complicated. 
There is an alternative ``collective'' encoding technique that allows $N$ qubits to be encoded in an ensemble of $N$ atoms, each of which has $N+1$ internal states $\ket{j}, ~j=0,N$ (\cite{Brion2007d}). 
The $j=0$ state is 
 a reservoir state, initially populated by all members of
the ensemble. The computational register with states $\ket{c_1,c_2,...c_N}$,  $c_j=(0,1)$
is represented by symmetric singly excited states of the $N$ atom ensemble with $c_j$ atoms in state $\ket{j}$ for $j>0$. The binary representation of a
register state thus corresponds to zero and unity populations of different internal states, which is enforced by Rydberg blockade across the ensemble of atoms (see Sec. \ref{sec.Rydberg}).  Only $N$ atoms is sufficient to realize this encoding, although a larger number may be preferable. A complete set of logical operations can be implemented without single particle addressing, as well as a form of hardware efficient error correction (\cite{Brion2008}).
This approach could potentially be scaled to large numbers of qubits in an array of Ho atom ensembles (\cite{Saffman2008}).

\subsection{Magic trapping}
\label{sec.magic}

Qubit encoding requires  the use of two or more internal states. Uncontrolled fluctuations in the differential trapping potential  dephase the qubit, so the coherence can be extended if the trapping potential is the same for both states. 
Traps for which this condition is satisfied are  referred to as ``magic traps'' (\cite{Ye2008}). The magic condition suppresses decoherence due to fluctuations in the trap laser power, or motion of the atoms within the trapping potential, thereby providing significant improvements in qubit coherence for both atomic clock and quantum computing applications. 

The design and practical implementation of magic trapping conditions is somewhat different for alkali and AEL atoms. For alkali 
atom clock qubits (states with $m_f=0$) the differential light shift (DLS) is approximated by Eq. (\ref{eq.DLS}) and cannot be canceled at large detunings which are required for low scattering rates and long coherence times. This conclusion has been verified by detailed calculations that include third order contributions to the polarizability arising from the combined effect of the quadratic ac Stark effect and the hyperfine interaction (\cite{Rosenbusch2009}). Remarkably, additional calculations that account for shifts at higher order than quadratic in the optical field strength reveal intensity magic conditions for clock state qubits (\cite{Carr2016}). It was also shown in the same work that by using bichromatic trap light with clock state qubits it is possible to achieve double magic conditions where both the DLS and the differential shift due to magnetic fields is first order insensitive to field variations. It is even possible to achieve ``triply'' magic trapping that is first order insensitive to intensity, frequency, and magnetic noise by taking advantage of both one- and two-photon contributions to the DLS (\cite{GLi2019}).   

For AEL atoms mitigation of DLS effects depends on the encoding and the specific isotope. The bosonic isotopes $^{88}$Sr and $^{174}$Yb have $I=0$ and no tensor shifts. The fermionic isotope $^{171}$Yb has $I=1/2$ and no tensor shifts, whereas the fermionic $^{87}$Sr has $I=9/2$ and extremely small tensor shifts (\cite{Westergaard2011}).  For  ground and metastable encodings in nuclear spin states the qubit coherence is essentially unperturbed by trap light intensity noise. For optical qubit encodings there is strong sensitivity to the DLS of the ground and metastable states. 
Long coherence times rely on the use of ``magic" wavelengths for which the ground and metastable states have the same polarizability, i.e.  $\alpha_{\rm g}(\omega)=\alpha_{\rm m}(\omega)$. This condition is only attained at a few discrete  wavelengths (\cite{Dzuba2010}). Summaries of magic wavelengths for Sr and Yb atoms 
are available in reference works (\cite{Pucher2026,Kroeze2026}).

Two-qubit entanglement via Rydberg interactions (see Sec. \ref{sec.Rydberg}) requires populating Rydberg states for a short time.ility of Rydberg states is usually very different from that of ground or metastable states, the trap light is typically tuned off during Rydberg gate operations. Since Rydberg gates are very fast turning off the traps for a short time does not cause excessive loss of atoms. Nevertheless, it is of interest to develop traps that can be kept on during Rydberg gate sequences.   The polarizability of an atom in a Rydberg state is essentially the ponderomotive potential of the Rydberg excited electron, weighted by the overlap of the Rydberg wavefunction with the trapping intensity (\cite{Dutta2000}). Hence Eq. (\ref{eq.alpha}) is modified to 
\be
U_{\rm R}(\omega)=-\frac{1}{4}\alpha(\omega) \int d^3r\, |\mcE({\bf R}+{\bf r})|^2 |\psi_{\rm R}({\bf r})|^2
\label{eq.alpharyd}
\ee
where $\psi_{\rm R}$ is the wavefunction of the Rydberg electron, which is assumed independent of the atomic center of mass position $\bf R$, $\bf r$ is the position of the electron relative to the center of mass, and the ponderomotive polarizability of the electron is $\alpha(\omega)=-e^2/{m_e\omega^2}$ with $e$ the electronic charge, and $m_e$ the electron mass.  . 

The optical polarizability of the Rydberg electron is negative, except near resonances with lower lying states, so tweezer traps act repulsively on Rydberg excited electrons. Bottle traps that require negative polarizability for atom confinement can be used to trap both ground and Rydberg states of alkali atoms (\cite{Barredo2020}).  Since the Rydberg wavefunction is delocalized compared to that of the ground state atom, achieving a magic condition between ground and Rydberg states requires appropriately sizing the trap intensity distribution. If this is done a ground-Rydberg magic condition can be achieved for alkali atoms over a wide range of wavelengths (\cite{SZhang2011}). 
 The situation is different for AEL atoms as the 2$^{\rm nd}$  valence electron which is not Rydberg excited can provide a strong and positive polarizability. If this  outweighs the negative polarizability of the Rydberg excited electron the atom can be trapped. Thus AEL atoms can be trapped in red detuned tweezers  as has been demonstrated with Yb atoms (\cite{Wilson2022}).

\subsection{Qubit  coherence}

Long coherence times that exceed qubit control and measurement times by orders of magnitude are a requirement for scalable quantum computing. Optically trapped neutral atom qubits have demonstrated coherence times that reach many seconds. 
Given quantum state control times of one $\mu\rm s$ or less, and measurement times of less than 1 ms, the available  coherence times provide a robust setting for implementing long quantum calculations.

Qubit coherence can be quantified in terms of $T_1, T_2^*,$ and 
$T_2$ times. The $T_1$ time which characterizes the rate of qubit bit flips is primarily limited by Raman scattering of the trapping light, and  can readily exceed tens of seconds in  far detuned tweezer traps or bottle traps, with a record $T_1=119 ~\rm s$ reported in an array of Cs atoms (\cite{Manetsch2025}). 
The qubit dephasing rate including both reversible and irreversible noise sources is characterized by the $T_2^*$ time. For alkali atoms the primary contributions to $T_2^*$ are magnetic noise and time varying differential light shifts from the trapping light, due either to laser noise or finite temperature atom motion. Reported coherence times include  
$T_2^*=44.9~\rm ms$ for Rb atoms in tweezer traps (\cite{Miles2026}),
$T_2^*=27.3~\rm ms$ for Cs atoms in tweezer traps (\cite{Miles2026}),
and 
$T_2^*=43~\rm ms$ for Cs atoms in bottle traps (\cite{Li2012}).

For AEL atoms 
encoded in ground or metastable nuclear spin states 
both magnetic and differential light shift noise are strongly suppressed giving long coherence times with $T_2^*=3.7~\rm s $ reported in $^{171}$Yb (\cite{Jenkins2022}).  With optical qubit encoding  good coherence is only obtained with magic trapping techniques (see Sec. \ref{sec.magic}) with which  
$T_2^*=3.4~\rm s$ has been demonstrated in  $^{88}$Sr atoms (\cite{Norcia2019}).  For the fine structure $^{88}$Sr qubit shorter coherence times of $T_2^*=2~\rm ms$ have been reported (\cite{Pucher2024}). 

The above results show that  $T_2^*$ coherence times in AEL atoms are almost 100 times longer than in alkali atoms. For both types of atoms the coherence can be improved further by applying dynamical decoupling pulses to cancel reversible dephasing (\cite{Suter2016}). This has led to coherence times of
$T_2=1-2 ~\rm s$ with Rb atoms in tweezer traps (\cite{Bluvstein2026}), 
$T_2= 12.6~\rm s$ with Cs atoms in tweezer traps (\cite{Manetsch2025}),
$T_2= 16.6~\rm s$ with Cs atoms in bottle traps (\cite{ZTian2024}),
and 
$T_2=40~\rm s$ in $^{87}$Sr  atoms
using ground nuclear spin encoding (\cite{Barnes2022}).
The $^{88}$Sr fine-structure qubit also benefits from dynamical decoupling, although in this case the  coherence achieved so far is  shorter than for the other encodings with $T_2=346~\rm ms $ the best result to date (\cite{Ammenwerth2025b}). 

With dynamical decoupling both alkali atoms and some encodings in AEL atoms  have reached coherence times longer than 10 s. Compared to measurement times of ms and sub ms (see Sec. \ref{sec.measurement}) these results provide a path towards deep, error corrected quantum computation with the ratio of coherence time to measurement time greater than $10^4$.  It is noteworthy that even without decoupling pulses, coherence times of several seconds have been achieved in AEL atoms. Thus these atoms may provide in some regards a simplified approach, with less control overhead, for reaching error corrected quantum computation.  

\subsection{Qubit initialization and measurement}

\label{sec.measurement}

Initialization and measurement are non-unitary operations that are necessary capabilities for quantum computation. 
Qubit initialization is readily performed by optical pumping techniques. While the specifics depend on the atom and choice of encoding, the general idea is that leveraging a combination of selection 
rules that depend on optical polarization and spectral selectivity, spontaneous decay from excited states can be used to pump atoms into states that are dark with respect to the pumping light. Doing so prepares well-defined fiducial states from which a computation can start. 

For alkali atoms pumping is readily performed into $m_f=0$ clock states using $\pi$ polarized light coupling to the lowest $p_{1/2}$ or $p_{3/2}$ levels, or into  stretched states with $m_f=f$ using $\sigma$ polarized light. In the latter case the atom is then  coherently transferred to a qubit clock state with a sequence of pulses. For AEL atoms pumping into a nuclear spin state with $M_F=\pm F$ is accomplished by driving the $^1{\rm S_0} - {^3}P_1$ transition with $\sigma_\pm$ polarized light. State preparation with fidelity above 0.999 is routine and detailed discussions of optical pumping techniques  can be found in standard references such as \cite{Auzinsh2010}.

Fast and high fidelity qubit measurements represent a more challenging task than state preparation. The standard approach to state measurement mirrors optical pumping in that a  combination of selection rules and spectral selectivity are used to render one qubit state bright and one qubit state dark to the measurement light. Nevertheless 
the time needed for a state measurement is generally much longer than the time needed for state preparation. The reason is that for state preparation every scattered photon contributes to the preparation process. Conversely for state measurement only the scattered photons that are collected by a lens and detected provide useful information. 
Even with fast imaging optics and sensitive detectors the efficiency with which a scattered photon is recorded is typically well under 10\%. Thus for the same scattering rates measurement may be more than ten times slower than state preparation. Speeding up measurement times is  important in order to establish a ratio between qubit coherence times and the time for a cycle of quantum error correction, that is large enough to enable efficient suppression of logical errors (see Sec. \ref{sec.QEC}). 

The number of scattered photons required to achieve a desired measurement fidelity can be estimated as follows (\cite{Saffman2005a}).  Assume that the mean rate of photon counts recorded by the measurement system  when the atom is in the ``dark'' state $\ket{0}$ is the background rate $r_{\rm b} ~(\rm s^{-1}).$ The mean rate of photon counts when the atom is in the ``bright'' state $\ket{1}$ is $r_{\rm b}+r_1 ~(\rm s^{-1})$ where $r_{\rm b}$ come from the background, which is the same for both states, and $r_1$ comes from the atom in the bright state.  
The processes responsible for the background rate $r_{\rm b}$ and bright state  rate $r_1$ are independent. The counting statistics for each process are described by Poisson distributions, i.e. the probability of recording $n$ counts in a time interval $t$ due to a  process with mean rate $r$ is 
$$
P_n(r,t)=\frac{(r t)^n e^{- rt}}{n!}.
$$ 
We set a cutoff value $n_{\rm c}$ and if the number of counts is $\ge n_{\rm c}$ the measurement result is $\ket{1}$ whereas if the number of counts is $<n_{\rm c}$ the measurement result is $\ket{0}$.

\begin{figure}[t!]
\centering\includegraphics[width=\textwidth]{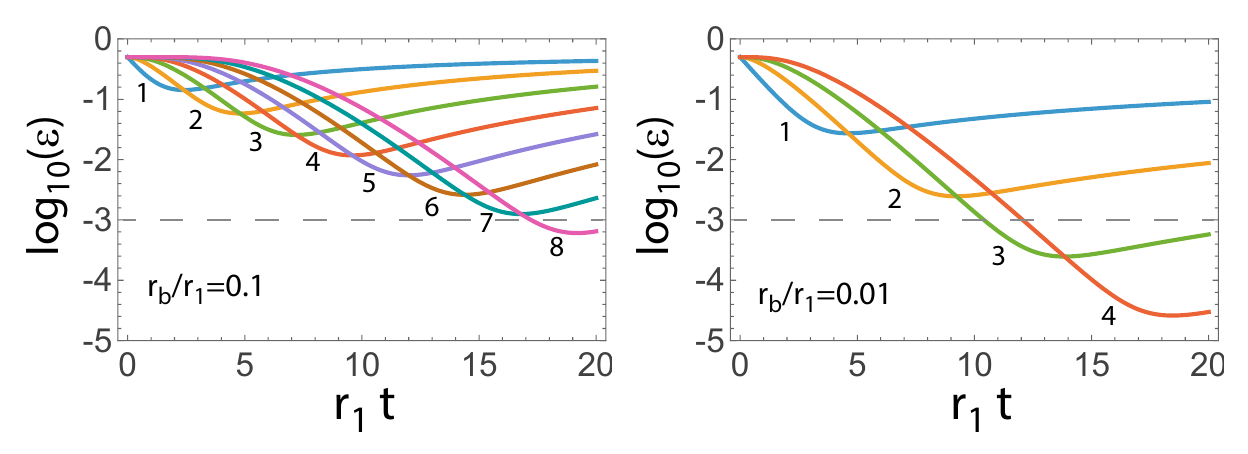}
\caption{Qubit state measurement error for a high background case $r_{\rm b}/r_1=0.1$ (left) and a low background case $r_{\rm b}/r_1=0.01$ (right). The plots are for $|c_0|^2=|c_1|^2=1/2$ and the curves are labeled with $n_{\rm c}$. The mean number of detector counts for an atom in the bright state is $r_1 t$.  }
\label{fig.measure}
\end{figure}

 If the atom is in state $\ket{0}$ the count distribution is due only to the background and is
$
P_{n,\ket{0}}=P_n(r_{\rm b},t)
%=\frac{(r_{\rm b} t)^n e^{- r_{\rm b}t}}{n!}
$
and the probability of an error is 
$$
\epsilon_{\ket{0}} = \sum_{n=n_{\rm c}}^\infty  P_{n,\ket{0}}.
$$
%This depends only on $r_{\rm b} t$ and $n_{\rm c}$. 
If the atom is in state $\ket{1}$ the count distribution is due to the independent background and signal counts giving
$
P_{n,\ket{1}}=\sum_{n_s=0}^n P_{n_s}(r_1,t)P_{n-n_s}(r_{\rm b},t)   
% =\sum_{n_s=0}^n \frac{(r_1 t)^{n_s} e^{- r_1t}}{n_s!} \frac{(r_{\rm b} t)^{n-n_s} e^{- r_{\rm b}t}}{(n-n_s)!}.
$  
and the probability of an error is 
$$
\epsilon_{\ket{1}} = \sum_{n=0}^{n_{\rm c}-1}  P_{n,\ket{1}}.
%= \sum_{n=0}^{n_{\rm c}-1}\sum_{n_s=0}^n \frac{(r_1 t)^{n_s} e^{- r_1t}}{n_s!} \frac{(r_{\rm b} t)^{n-n_s} e^{- r_{\rm b}t}}{(n-n_s)!}.
$$
%This depends  on $r_{\rm b} t, r_1 t$ and $n_{\rm c}$. 
Thus the probability of a measurement error for an arbitrary state $\ket{\psi}=c_0\ket{0}+c_1\ket{1}$  is 
\bea
\epsilon &=& |c_0|^2 \epsilon_{\ket{0}} + |c_1|^2 \epsilon_{\ket{1}} \nn\\
&=&|c_0|^2\left[  1-\frac{\Gamma(n_{\rm c},r_{\rm b} t)}{\Gamma(n_{\rm c})}\right] + |c_1|^2 \frac{\Gamma(n_{\rm c},(r_1+r_{\rm b})t)}{\Gamma(n_{\rm c})}.\label{eq.qubiterror2}
\eea
Here $\Gamma(n)=\int_0^\infty dt\, t^{n-1} e^{-t}=(n-1)!$ for integer $n$ is the gamma function and $\Gamma(n,x)=\int_x^\infty dt\, t^{n-1} e^{-t}$ is the incomplete gamma function. 
Figure \ref{fig.measure} shows representative results for several values of $r_{\rm b}/r_1$ and $n_{\rm c}$. For the high background case an error of 0.001 requires 
$r_1 t > 17$ and for the low background case $r_1 t > 10$. In the limit of zero background 
$\epsilon=|c_1|^2 e^{- r_1 t}.$
Reaching a target high fidelity of $\epsilon=0.0001$ for the uniform superposition state requires $r_1 t=8.5$ detected photons. 

We can use this last result to estimate the minimum possible measurement time based on light scattering. Assuming an optimistic 10\% efficiency including photon collection, optical losses, and detector quantum efficiency the number of scattered photons must be at least 85 to reach $\epsilon=0.0001$. Different possible scattering transitions and the corresponding minimum times to scatter say, 100 photons are listed in Table \ref{tab.scattering}. For alkali atoms 
 scattering on
the $\ket{n_g s_{1/2},f=I+1/2} \leftrightarrow \ket{n_g p_{3/2},f'=I+3/2}$ cycling transition,  $t_{100}\sim 5~\mu\rm s$. Fast scattering on this transition is only viable without hyperfine changing Raman transitions by using $\sigma$ polarized readout light in a 1D geometry (\cite{Martinez-Dorantes2017,Kwon2017}).  
 For AEL atoms even faster scattering is possible using the strong $\ket{^1{\rm S_0}}  \leftrightarrow \ket{^1{\rm P_1}}$  transition for which 
 $t_{100}\sim 1~\mu\rm s$. This is possible in $^{171}$Yb but $^{87}$Sr
 has a leak from $^1{\rm P_1}$ to a lower lying $^1D_2$ state which  impedes reaching $\epsilon=0.0001$ without additional repumpers.  AEL atoms can also be measured by cycling on the narrow  $\ket{^1{\rm S_0}}  \leftrightarrow \ket{^3{\rm P_1}}$ transition. In $^{87}$Sr this is very slow but could be as fast as 175 $\mu\rm s$ for $^{171}$Yb. Finally it is possible to cycle on $^3{\rm P_2} \leftrightarrow {^3}D_3$ transitions between excited states in AEL atoms, although the leakage properties of these transitions have not yet been carefully validated.   

 The measurement time estimates given in Table \ref{tab.scattering} are informative, but not realistic without accounting for the heating and possible loss of atoms that accompany high rate scattering. In Cs atoms a novel approach can be based on cycling via  
 $$
 \ket{6s_{1/2},f=4} \xrightarrow{685~\rm nm} \ket{5d_{5/2},f=6}
 \xrightarrow{3491~\rm nm}  \ket{6p_{3/2},f=5}
 \xrightarrow{852~\rm nm}   \ket{6s_{1/2},f=4}. 
 $$ 
 In this method the excitation at 685 nm is background free for 852 nm detection. Furthermore, low leakage cycling can be combined with 3D cooling. The measurement speed 
is not intrinsically fast due to the narrow  linewidth of $5d_{5/2}$
but can be sped up by active quenching of the excited state (\cite{Scott2025}). Progress on fast measurements in AEL atoms has been made based on rapidly strobing counterpropagating beams in a high saturation regime to minimize momentum transfer (\cite{Bergschneider2018,LSu2025}). With this approach imaging of single $^{171}$Yb atoms with fidelity 0.999 and retention probability 0.995 was achieved in 6.4 $\mu\rm s$ (\cite{Falconi2025}).  The effective measurement time including recooling was several hundred $\mu\rm s$.  
The heating can be reduced using a coherent excitation scheme where a sequence of deterministic $\pi$ pulses is used to periodically excite the atoms (\cite{Yokoyama2026}). This suppresses momentum fluctuations due to the absorption process, leaving only momentum transfer and heating from photon emission.

\begin{table}[!t]
\centering
\caption{Maximum resonant scattering rates for common laser-cooling transitions. The maximum rate is
$r_{\rm max}$ for a closed, saturated transition. The last column is the time to scatter 100 photons. For Cs readout via $5d_{5/2}$ the excitation is at 685 nm and readout is at 852 nm (\cite{Scott2025}).}
\begin{tabular}{|l|l|c|c|c|c|}
\hline
Species & Transition & $\lambda$ (nm) & $\Gamma/2\pi$ (MHz) & $r_{\rm max}$ (photons/$\mu$s)& $t_{100}$  ($\mu\rm s$) \\
\hline
$^{87}$Rb & 5s$_{1/2}\rightarrow5$p$_{3/2}$ & 780 & 6.07 & 19.1& 5.24 \\
$^{133}$Cs & 6s$_{1/2}\rightarrow6$p$_{3/2}$ & 852 & 5.23  & 16.4 & 6.10\\
$^{133}$Cs & 6s$_{1/2}\rightarrow5$d$_{5/2}$ & (685)852 & 0.118  & 0.371 & 270.\\
$^{87}$Sr & $^1$S$_0\rightarrow{}^1$P$_1$ & 461 & 30.5 & 95.9 & 1.04\\
$^{87}$Sr & $^1$S$_0\rightarrow{}^3$P$_1$ & 689 & 0.0074  & 0.023 & 4350.\\
$^{171}$Yb & $^1$S$_0\rightarrow{}^1$P$_1$ & 399 & 28.9 & 90.8 & 1.10\\
$^{171}$Yb & $^1$S$_0\rightarrow{}^3$P$_1$ & 556 & 0.182 & 0.572 & 175.\\
\hline
\end{tabular}
\label{tab.scattering}
\end{table}

\section{Single qubit gates}

Single qubit gate operations can be implemented with microwave or optical frequency fields. A first example is provided by a qubit encoded in hyperfine-Zeeman clock states of an alkali atom $\ket{0}\equiv\ket{f_-,0}$, $\ket{1}\equiv\ket{f_+,0}$ where $f_\pm=I\pm 1/2$ with $I$ the nuclear spin. With a weak magnetic bias field to stabilize the $m_f=0$ state against precession towards other $m_f\ne 0$ states the clock states can have lifetimes of many seconds. Although these states are coupled by a magnetic dipole transition the intrinsic rate of state changing events in a room temperature environment is extremely small, about once every 35 years for Cs atoms. The rate can be reduced further by operating in a cryogenic environment, which has additional advantages in terms of lower background pressure (\cite{Schymik2021}) and longer Rydberg state lifetimes (\cite{JJIn2026}). 

%There are three common methods for controlling the quantum state of a hyperfine encoded qubit: microwaves, light shift gates, and Raman gates.

Microwaves resonant with the clock transition frequency (6.8 GHz for $^{87}$Rb or 9.2 GHz for $^{133}$Cs) can be used to simultaneously apply $\sf R(\theta,\phi)$ gates to  atoms in large arrays. Here $\theta$ is the pulse area while the axis of rotation lies in the equatorial plane of the Bloch sphere, and makes an angle of  $\phi$ with respect to the $x$ axis. The  pulse area is controlled by adjusting the duration or amplitude of the microwave pulse while $\phi$ is varied by changing the phase of the microwave signal relative to the system phase reference.

A single  $\sf R(\theta,\phi)$ gate  has the matrix representation in the qubit basis 
\begin{equation}
    {\sf R}(\theta,\phi) =
    \begin{pmatrix}
    \cos(\theta/2) & -i e^{-i \phi}\sin(\theta/2) \\
-i e^{i \phi}\sin(\theta/2) & \cos(\theta/2)
\end{pmatrix}.
\end{equation}
Concatenating three  rotations gives an arbitrary unitary $\sf U$ using the decomposition
\begin{equation}
{\sf U}={\sf R}(\gamma,0)
{\sf R}(\beta,\pi/2){\sf R}(\alpha,0).
\end{equation}
Only two phases $0$ and $\pi/2$ are required and explicit expressions for the pulse areas $\alpha,\beta,\gamma$ in terms of the elements of $\sf U$ can be found in standard texts (\cite{Nielsen2000}).

Microwave driven gates in 2D arrays were first demonstrated in \cite{Xia2015} and can reach fidelities above 0.9999 (\cite{Nikolov2023}). 
While microwave gates are relatively straightforward to implement they do not provide single qubit addressability since the cm scale microwave wavelength is thousands of times longer than the few micron spacing in atom arrays. Single qubit gates on selected sites can be implemented using a combination of microwaves and local addressing with tightly focused laser beams (\cite{Xia2015}). This approach can  be extended  to site-selective control within 3D arrays (\cite{YWang2016}).

\begin{figure}[t!]
\centering
\includegraphics[width=.4\textwidth]{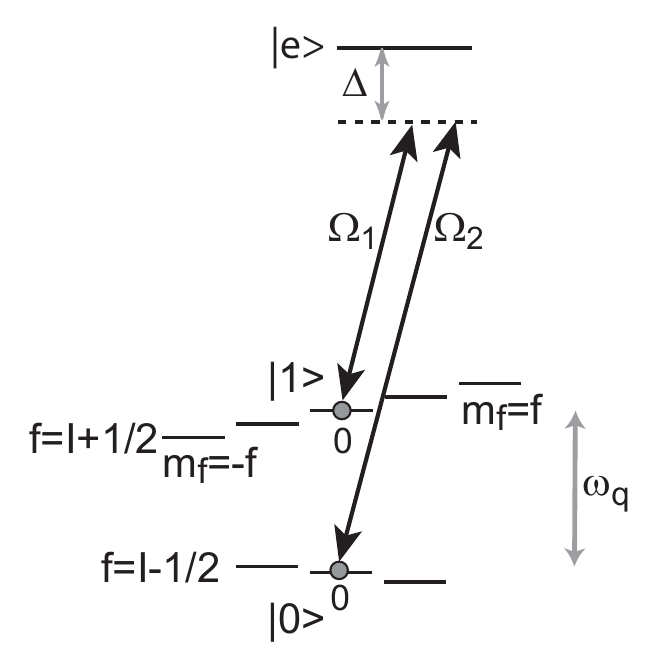}
\caption{One qubit gates in alkali atoms can be driven by microwaves at frequency $\omega_q$ or optical Raman transitions.   }
\label{fig.1qgate}
\end{figure}

A purely optical solution that directly enables site-selective control replaces the microwaves with an optical beam containing two components separated by the qubit frequency $\omega_{\rm q}$ as shown in Fig. \ref{fig.1qgate}. In terms of  one-photon Rabi frequencies $\Omega_1, \Omega_2$ the two-photon Rabi frequency connecting the qubit 
states is $\Omega=\Omega_1\Omega_2^*/(2\Delta)$ where $\Delta$ is the detuning from the excited state. Arbitrary ${\sf R}(\theta,\phi)$ rotations can be performed, with the phase $\phi$ now controlled by the relative phase of the one-photon fields. These optical gates can be driven at higher rates than microwave gates, with several MHz 
possible with moderate optical power  (\cite{Knoernschild2010}). This type of gate is subject to  scattering via partial excitation of the excited state. The scattering error scales as (\cite{Saffman2005a}) 
$\epsilon\sim \Gamma/\Delta$  with $\Gamma$ the decay rate of the excited state. 
With optimized parameters scattering errors of less than
$2\times 10^{-4}$ per $\pi$ pulse have been demonstrated (\cite{Levine2022}).

The discussion so far in this section is relevant for qubits encoded in alkali atom hyperfine-Zeeman states.  
For AEL atoms the {\it omg} architecture affords several different encoding options, each of which requires specific approaches for single qubit gate operations. 
One encoding choice that can be used for $^{87}$Sr $(I=9/2)$ or $^{171}$Yb $(I=1/2)$ is to use nuclear spin states $\ket{0}=\ket{^1{\rm S_0},M_F}$, $\ket{1}=\ket{^1{\rm S_0},M_F+1}$  in the ${^1}{\rm S_0}$ electronic ground state (see Fig. \ref{fig.encoding}). In this situation the qubit states are energy degenerate in the absence of a magnetic field and experience a purely linear Zeeman splitting for moderate magnetic fields since with $J=0$  there is no hyperfine coupling giving nonlinear mixing of hyperfine levels.  For  $^{87}$Sr neighboring states are split by  185 Hz/G and for $^{171}$Yb 750 Hz/G. 

The small splitting for Gauss level fields renders these qubits naturally insensitive to magnetic noise which leads to coherence times of many seconds (\cite{Barnes2022}). However, single qubit gates are typically slower than in alkali atoms. For $^{87}$Sr with $I=9/2$ it is not possible to drive single photon radio frequency   rotations between neighboring $M_F$ states without leakage to other states outside of the computational basis since each pair of adjacent states has the same energy difference. For
$^{171}$Yb with $I=1/2$ there are only two states with $M_F=\pm 1/2$ and radio frequency fields can be used to drive rotations on qubits encoded in ground or metastable nuclear spin states, in the same way that microwaves are used for alkali atoms. However, the achieved rates with radio frequency driving have so far been limited  to less than 1 kHz (\cite{Ma2022, SMa2023}). 

Faster rates are possible with optical Raman driving. 
For $^{87}$Sr this has been done
with one $\pi$ polarized component and one $\sigma$ polarized component. To prevent leakage out of the computational basis the neighboring $M_F$ state is Stark shifted out of resonance with the Raman drive using an additional optical frequency component. This approach has yielded Rabi frequencies of 1.16 kHz  (\cite{Barnes2022}).
Notable results for $^{171}$Yb are Rabi frequencies of 7 kHz with Clifford fidelities of 99.963\%  (\cite{Muniz2025a})
and Rabi frequencies of 1.47 MHz with Clifford fidelities of 99.48\%  (\cite{Jenkins2022}).

The other {\it omg} encoding option is an optical frequency qubit with
the basis states $\ket{0}=\ket{^1{\rm S_0},M_F}$, $\ket{1}=\ket{^3{\rm P_0},M_F}$. Qubit rotations can be driven with one-photon optical transitions. As this is an intercombination transition between singlet and triplet states the matrix element is small and for moderate optical intensities Rabi rates of 110 kHz with Clifford fidelity of 99.804\% have been achieved in  $^{171}$Yb (\cite{Lis2023}).

\section{Entangling gates}

In the late 1990s and early 2000s many proposals were put forward for quantum logic with neutral atom qubits. These included 
photon mediated interactions in optical cavities (\cite{Pellizzari1995}),
atomic collisions (\cite{Jaksch1999}),
short range dipole enhanced interactions (\cite{Brennen1999}), and Rydberg interactions (\cite{Jaksch2000}).
The most successful of these ideas have been the use of atomic collisions (\cite{Jaksch1999}) and Rydberg interactions (\cite{Jaksch2000,Lukin2001}). 

\subsection{Cold collisions}

The basic idea for atomic collisions is that the phase acquired during an atomic collision is dependent on the internal electronic states of the atoms which are used to encode qubits. Using state dependent optical potentials atoms can be moved in a controlled fashion to implement entangling collisions via  state dependent phase shifts on the wavefunction (\cite{Jaksch1999}). With atoms trapped in an optical lattice collisions can be implemented in parallel on many pairs, thereby entangling  many atom pairs in parallel. This type of approach was first demonstrated in optical lattices (\cite{Mandel2003,Anderlini2007}) and more recently was shown to reach very high entanglement fidelities near 99.9\% simultaneously across thousands of atom pairs (\cite{Bojovic2026,Kiefer2026}). The massive entanglement thereby obtained is of great interest for quantum simulation, but is not immediately applicable for digital quantum computation due to the lack of single site control. 

The requisite individual qubit control can be realized using atoms in movable optical tweezers. Two-qubit entanglement has been demonstrated in this way (\cite{Kaufman2015}), although the reported fidelity was substantially below that obtained in the recent lattice experiments.  It is also possible to consider hybrid approaches that combine the parallel operation afforded by lattices, with the individual control of tweezers (\cite{Weitenberg2011b}).  A challenge for collision mediated entanglement is that the collisional phase directly depends on the center of mass motional state of the atom, even though the quantum information is encoded in spin states. Cooling close to the motional ground state is therefore a prerequisite for high fidelity entanglement. 

\subsection{Rydberg interactions}
\label{sec.Rydberg}

\begin{figure}[!t]
\center
\includegraphics[width=.8\columnwidth]{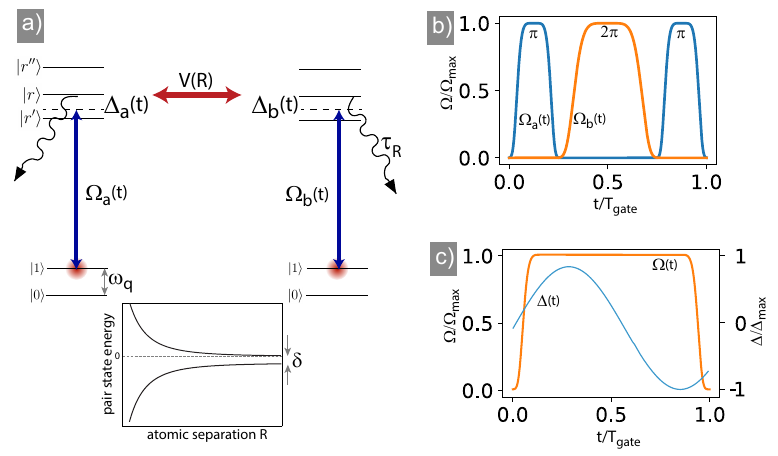}
\caption{\label{fig.gate} Rydberg interactions and gate protocols. a) Two qubits are coupled to Rydberg state $\ket{r}$ with Rabi rate $\Omega(t)$ and detuning $\Delta(t)$, which may be different for the two atoms. Neighboring opposite parity states $\ket{r'},\ket{r''}$ are not optically excited due to selection rules, but are coupled to $\ket{r}$ via dipole  allowed, microwave frequency transitions. The energy defect (sometimes called a F\"orster defect) is $\hbar\delta=U_{r''} + U_{r'}-2 U_r.$ The lower inset shows the pair state energies for a near-resonant case with $\delta\ne 0.$
b) The three pulse gate (\cite{Jaksch2000}) uses a resonant $\pi$ pulse on atom a, a $2\pi$ pulse on atom b, and a $\pi$ pulse on atom a.  c) The time-optimal gate (\cite{Jandura2022}) uses the same drive on both atoms with a constant Rabi frequency, and a time varying detuning that can be approximated by an offset sinusoid. }
\end{figure}

Rydberg interactions provide an alternative, and  widely adopted entanglement mechanism that has reduced sensitivity to the motional state. The original proposal for Rydberg mediated entanglement (\cite{Jaksch2000}) relies on the concept of Rydberg blockade (\cite{Lukin2001}) whereby excitation of a single atom to a Rydberg state prevents atoms less than a blockade distance away from also being excited. The basic scheme for entanglement generation is shown in Fig. \ref{fig.gate}.  Optical pulses couple qubit states $\ket{1}$ to a Rydberg level $\ket{r}$ with time dependent Rabi frequency $\Omega(t)$ and detuning $\Delta(t)$. The same pulses may be applied to both atoms, or different pulses may be applied to each atom. The energy difference \footnote{Here and in what follows we implicitly set $\hbar=1$ and describe energies in frequency units.} of the qubit states $\omega_{\rm q}$ is assumed much larger than $\Omega$ so that the $\ket{0}$ states are not Rydberg excited. 

Rydberg states are atomic orbitals with definite parity and therefore have no permanent dipole moment. The atom-atom interaction takes the form of an effective dipolar interaction $V_{\rm dd}\sim d'd''/R^3$, where $d'=\bra{r'}r\ket{r}$, $d''=\bra{r''}r\ket{r}$, with $r$ the electron coordinate relative to the atomic core (\cite{Walker2008}). When the energy defect vanishes, which can occur for some choices of states, the interaction is resonant and the pair population oscillates between $\ket{rr}$ and $ \ket{r'r''}$,  $\ket{r'',r'}$.
This interaction mechanism was introduced by F\"orster in the context of molecular energy transfer (\cite{Forster1948}). 
At large separations  the   energy defect $ \delta > V_{\rm dd}$ and we obtain a  second order virtual interaction with van der Waals type scaling $V_{\rm vdW}\sim V_{\rm dd}^2/{\delta }$.
The transition from the $1/R^3$ resonant regime to the $1/R^6$ van der Waals limit is smooth, and is evident in the interaction curves shown in Fig. \ref{fig.rydberg}. 
Note that the interaction does not involve matrix elements between different atoms. In this sense the interaction is always of a virtual nature. The implication is that different states $\ket{r}$ can be chosen for the two atoms, as long as the energy defect is small. The two atoms can even be chosen to be different elements, while still retaining a strong resonant interaction (\cite{Beterov2015,Anand2024}).

Many different pulse profiles and gate designs are possible (\cite{Saffman2010, XFShi2022}). The original three pulse design (\cite{Jaksch2000}) provides an easily understood example.  
In the ideal limit of strong  blockade  $V\gg \Omega$, $\omega_{\rm q}\gg \Omega$ and long Rydberg lifetime $\tau_{\rm R}\gg T_{\rm gate}$, with $T_{\rm gate}$ the duration of the Rydberg pulses the three pulse protocol shown in Fig \ref{fig.gate}b) leads to a controlled phase gate 
$$
{\sf CZ'}=\begin{pmatrix} 1 & 0 & 0 & 0\\0 & -e^{\imath\phi_{01}} & 0 & 0 \\ 0 & 0 & -e^{\imath\phi_{10}} & 0 \\ 0 & 0 & 0 & -e^{\imath(\phi_{01}+\phi_{10})} 
\end{pmatrix}.
$$
The gate matrix has been expressed in the computational basis $\{ \ket{00}, \ket{01},\ket{10},\ket{11}\}$. The  dynamical $\pi$ phase shifts on all but the first row are due to the excitation and de-excitation of an atom to the Rydberg state. Crucially, it is the strong Rydberg blockade interaction that prevents double excitation of the $\ket{11}$ state and ensures preparation of entanglement.  The additional phases $\phi_{01}$ and $\phi_{10}$ arise from single photon light shifts imparted by the Rydberg beams (\cite{Maller2015}). 
Application of local ${\sf R_Z}$ rotations leads, up to an irrelevant global phase, to a canonical phase gate ${\sf CZ}={\rm diag}(1,1,1,-1)$.

In practice neither the interaction strength $V$ nor the Rydberg lifetime $\tau_{\rm R}$ are arbitrarily large and the minimum entanglement error scales as $\epsilon= \eta /(V\tau_{\rm R})^{2/3}$ with $\eta=3 (7\pi)^{2/3}/8\approx 2.9$ (\cite{XZhang2012}). This minimum error is achieved by optimizing the Rabi frequency to balance the error due to imperfect blockade of the $\ket{11}$ state which scales as $(|\Omega|/V)^2$ against the error from Rydberg decay which scales as $1/(|\Omega|\tau_{\rm R})$.  The $1/(V\tau_{\rm R})^{2/3}$ error scaling turns out to be an artifact of the three pulse protocol and an analysis for arbitrary pulses shows that the fundamental error floor for Rydberg entanglement is (\cite{Wesenberg2007}) 
\begin{equation}
\epsilon= \eta/(V\tau_{\rm R}).
\label{eq.epsmin}
\end{equation}
This scaling is found by evaluating the dynamical evolution of the Schmidt coefficients that determine the amount of entanglement and finding the minimum  integrated  
population of the Rydberg states required for producing one unit of entanglement. The analysis provided a bound of $\eta\ge 2.09$ which was later made tight, and shown to be $\eta_{\rm min}=1+\pi/2 \approx 2.57$ (\cite{Doultsinos2025b}).  The bound can be further improved with rank 2 driving that  excites both $\ket{0}$ and $\ket{1}$ states to different Rydberg levels in the presence of a F\"orster resonance, in which case $\eta_{\rm min}=\pi/2\approx 1.57$ (\cite{Norrell2026}).

Although Rydberg blockade provides an intuitive explanation for the operation of the three pulse gate, extremely strong blockade is not a requirement for high fidelity entanglement. This was pointed out in the original gate paper (\cite{Jaksch2000}) where a protocol based on the interaction of pairs of rapidly excited Rydberg atoms in the limit $|\Omega| > V$ was described. In this interaction mode of operation, entanglement requires $V T=\pi$, where $T$ is the interaction time. This implies a gate error $\epsilon=T/\tau_{\rm R} = \pi/(V\tau_{\rm R})$ which has $\eta=\pi$. Although this $\eta$ is not excessively  large compared to  the best possible value with rank 1 driving of $\eta_{\rm min}=1+\pi/2$ (\cite{Doultsinos2025b}) it is preferable not to have both atoms fully Rydberg excited which leads to interatomic forces and motional heating.

In order to minimize the gate error when $V > |\Omega|$ so double excitation is minimized, the pulse has to be carefully designed. The challenge is that single atom excitation proceeds with Rabi frequency $\Omega$ whereas two 
atom excitation proceeds with Rabi frequency $\sqrt{2}\Omega$ when there is strong blockade. 
The reason is that in the limit of strong blockade the $N$ atom state $\ket{\bar 1}=\ket{1_1 1_2 ... 1_N}$ couples to the singly excited Rydberg superposition state
$\ket{\bar r}=(1/N^{1/2})\sum_{j=1}^N \ket{1_1 1_2 ... r_j ...1_N}$. 
Each of the $N$ kets in the superposition couples to the ground state at rate $\Omega/N^{1/2}$ and since there are $N$ terms the interaction between $\ket{\bar 1}$ and $\ket{\bar r}$ has an effective coupling rate $\Omega_N=N^{1/2} \Omega.$ 
The $N^{1/2}$ scaling has been verified  in several experiments (\cite{ Dudin2012b,Ebert2014,Zeiher2015}).

The $N^{1/2}$ scaling implies that a pulse with constant amplitude and phase that has a pulse area of $\pi$ for a single atom will have a pulse area of $\sqrt2 \pi$ for two atoms leading to large gate errors. This can be solved using smoothly varying adiabatic pulses, but the time spent in the Rydberg state is longer than desired (\cite{YSun2020,Saffman2020}). The optimal approach, as shown in Fig. \ref{fig.gate}c), turns out to be based on simultaneously applying to both atoms a constant amplitude Rabi drive with a time varying detuning. The pulse is designed such that even though the one- and two-atom cases have different effective Rabi frequencies, in both cases the atomic population returns to the ground state with the correct dynamical phases for  entanglement (\cite{Jandura2022, Pagano2022, Mohan2023}).

High fidelity is achieved with time optimal pulses  when the pulse duration $T_{\rm gate}$ satisfies (\cite{Jandura2022})  $T_{\rm gate}\ge 7.6/|\Omega|$, with $\Omega$ the ground - Rydberg Rabi frequency. The time-optimal pulse design eliminates the $(|\Omega|/V)^2$ coherent error characteristic of the three pulse gate. The gate error is then determined by the integrated population in the Rydberg state which is approximately $T_{\rm R}\ge 2.96/|\Omega|$ which leads to 
\begin{equation}
\epsilon_{\rm min}^{\rm (TO)}=\frac{T_{\rm R}}{\tau_{\rm R}}\approx
\frac{2.96}{|\Omega|\tau_{\rm R}}.
\end{equation}
In the strong blockade limit of $|\Omega|\ll V$ this error is worse than the fundamental bound of Eq. (\ref{eq.epsmin}). However time-optimal type pulses can be designed for $|\Omega|\sim V$ or even $|\Omega|> V$  in which case the gate fidelity approaches the fundamental limit (\cite{Poole2025a}).

The  fidelity of a Rydberg gate in a real atom depends on the atomic interaction strength and excited state lifetime. The dependence on atomic separation is shown in Fig. \ref{fig.rydberg}. For alkali atoms Rydberg interaction parameters for specific states and geometries can be  accurately calculated using tabulated quantum defect values (\cite{Walker2008}). For AEL atoms with two valence electrons detailed calculations are more complex and require multi-channel quantum defect theory for accurate results (\cite{Peper2025}).

\begin{figure}[!t]
\center
\includegraphics[width=.7\columnwidth]{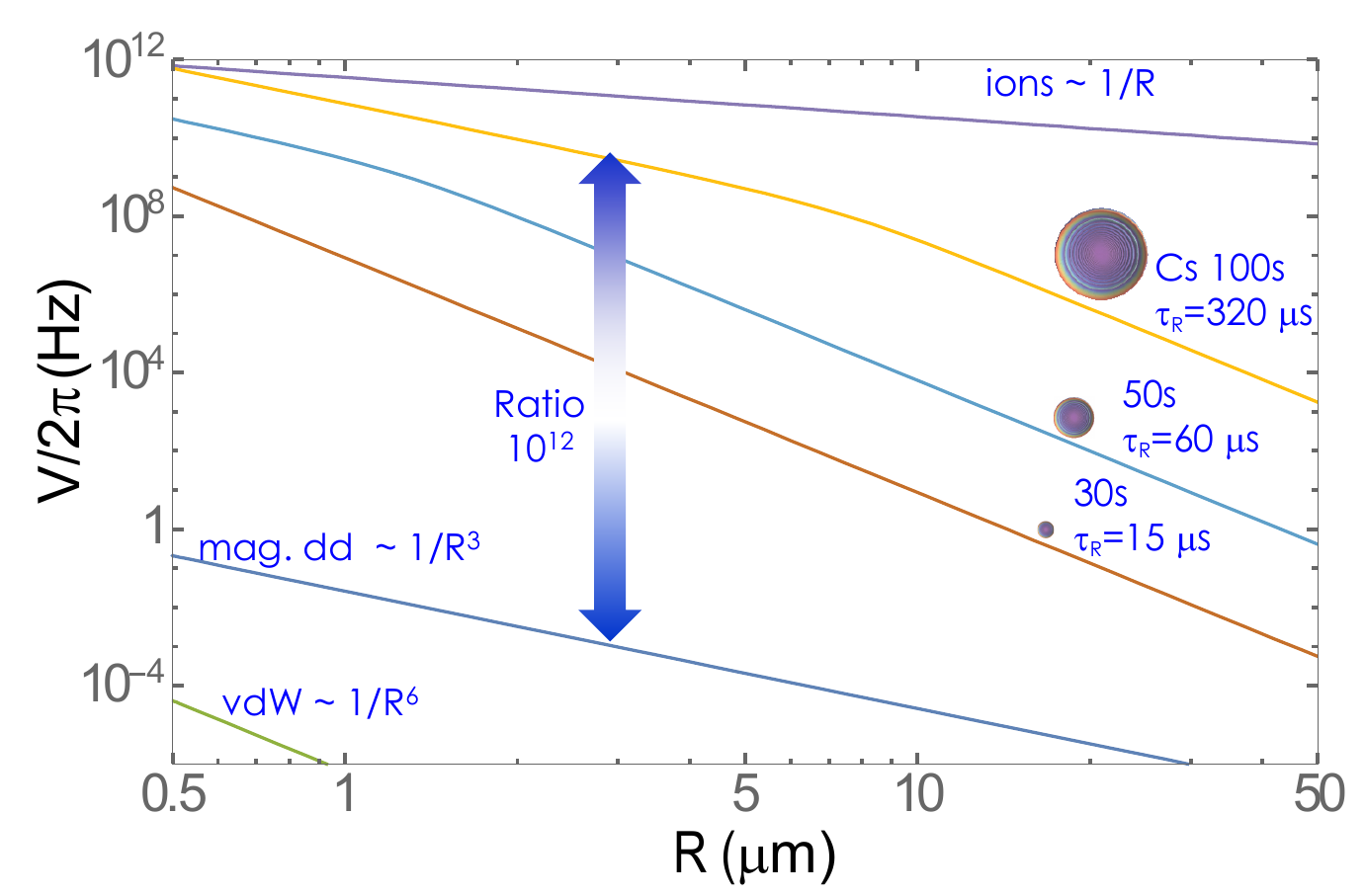}
\caption{\label{fig.rydberg} Interaction energy of ground state and Rydberg excited Cs atoms as a function of the interatomic spacing $R$. The Rydberg curves are labeled with the room temperature radiative lifetimes, including blackbody induced depopulation (\cite{Beterov2009}). The curve labeled ions is the Coulomb interaction of singly charged ions. }
\end{figure}

For Cs 100s states, as shown in Fig. \ref{fig.rydberg}, and using  $V/|\Omega|=10$ at $R=10(20)~\mu\rm m$ the time-optimal minimum error is 
$$
\epsilon_{\rm min}^{\rm (TO)}=0.00029 ~(0.016). 
$$
The rank 1 drive minimal error for the same conditions is
$$
\epsilon_{\rm min}=0.000026 ~ (0.0014). 
$$
We see that using optimized pulse designs high fidelity gates are possible at distances up to about $20~\mu\rm m$. At shorter distances of $2-3 ~\mu\rm m$ the interaction is substantially stronger and high fidelity is possible with lower Rydberg states near $n=50-60$  that are less sensitive to external electric fields (\cite{Evered2023,Tsai2025}). 

In recent years there has been remarkable experimental progress in implementing high fidelity Rydberg gates. The original demonstrations produced states not much over the threshold for entanglement (\cite{Wilk2010, Isenhower2010, Zhang2010}). These early demonstrations of a $\sf CNOT$ gate used the three pulse protocol. Developments that have led to much higher fidelity included better understanding and reduction of the deleterious effects of laser phase noise (\cite{Leseleuc2018,Levine2018,XJiang2023}) and the introduction of improved gate protocols, particularly the time-optimal protocol discussed above. The most recent results have achieved entanglement fidelity better than ${\mathcal F}=0.99$
with $^{87}$Rb (\cite{Evered2023}), $^{133}$Cs (\cite{Radnaev2025}), $^{88}$Sr (\cite{Tsai2025}) and $^{171}$Yb (\cite{BZhang2026}) atoms. These results are below the threshold needed for quantum error correction, but still far from the fundamental limits set by atomic physics. Further improvement will require  increasing the Rydberg Rabi rate to tens of MHz, reduction of laser noise, even colder atoms to reduce errors associated with atom motion and position variations, management of the effects of photon recoil (\cite{Robicheaux2021}),  and reduction of disturbances from background electromagnetic fields. Operation in a cryogenic environment can also help by increasing the Rydberg lifetime (\cite{JJIn2026}). There are also efforts to utilize the much longer lifetimes of circular Rydberg states in cryogenic environments, particularly  for quantum simulation (\cite{TLNguyen2018}).

\section{Quantum error correction}
\label{sec.QEC}

It is widely believed that realization of broad quantum computational advantage relative to conventional computers will require incorporation of quantum error correction (QEC) techniques (\cite{Terhal2015, Roffe2019}). QEC is a vast topic; here we  provide only a brief introduction emphasizing aspects that leverage the capabilities of atom based quantum computing.  
Qubit errors are more difficult to correct than bit flips in digital, binary computers, in part because  qubits are subject to continuous errors in both amplitude and phase, so there are more possible errors to correct. In addition, simple repetition codes which can be used to provide resilience against errors in conventional systems, are not possible in a quantum setting due to the no-cloning theorem (\cite{Wootters1982}). Furthermore, measuring qubits to determine if they have been corrupted collapses quantum superposition states and would interrupt a calculation. Despite these challenges it was shown early on that QEC is possible using ingenious encoding and measurement strategies that enable enough information to be gathered to correct errors, without collapsing the logical quantum state (\cite{Shor1995,Steane1996b}).

QEC codes are labeled with the notation $[[n,k,d]]$  where $n$ is the number of physical qubits, $k$ is the number of encoded logical qubits, and $d$ is the code distance. 
In addition to the $n$ physical qubits some number of ancilla qubits are needed whose role is not to store computational states, but which act as auxiliary qubits that are entangled with data qubits, and then measured in order to extract error information. Typically there are ${\mathcal O}(n)$ ancilla qubits.  
The code rate $r=k/n$ quantifies the efficiency of the encoding.  
For a single logical qubit, $k=1$, the minimum number of physical qubits that can correct an arbitrary single error is $n=5$, so the maximum code rate is $r=1/5.$ The limit $n=5$ can be deduced from a simple counting argument. The Hilbert space dimension of $n$ qubits is $2^n$ while the number of possible distinct errors for the initial states $\ket{0}_{\rm logical}$ and $\ket{1}_{\rm logical}$  is $2(3n+1)$, including the no error condition. Requiring that an error is detectable corresponds to $2^n\ge 2(3n+1)$, so $n\ge 5$.

The code distance determines how effectively logical errors are suppressed, with the probability of a logical error scaling as 
$$
p_{\rm logical}\sim (p/p_{\rm th})^{\frac{d+1}{2}},
$$
where $p$ is the physical error probability and $p_{\rm th}$ is the threshold error probability, which depends on the choice of code. Although a $[[5,1,3]]$ code with $n=5$ exists (\cite{Laflamme1996}) the corresponding value of $p_{\rm th}$, and the small $d=3$ code distance  are not favorable for suppressing errors in deep logical circuits.  Codes with larger values of $n$ provide greater code distance and exponentially better error suppression. A primary example is the widely studied surface code (\cite{Fowler2012}) which only requires nearest neighbor connectivity on a two-dimensional planar grid. These connectivity requirements make the surface code compatible with many different types of qubits at the expense of a code distance that scales as $d\sim \sqrt n$, implying a large overhead in physical qubit count to reach a high code distance. 

There have been several demonstrations of logical encoding with neutral atom qubits (\cite{Bluvstein2024,Muniz2025b,WChung2025,Bluvstein2026,BZhang2026}) as well as the preparation of resource states for performing universal logical gates (\cite{Rodriguez2025}). The largest system demonstrated to date used up to 96 logical qubits with $d=4$ and showed that scaling the surface code up to $d=5$ gave a below threshold logical performance factor of $\times 2.14$ improvement with a machine learning based circuit for decoding errors and detection of atom loss events  (\cite{Bluvstein2026}).

 A figure of merit  that captures both the encoding rate $r$ and the code distance $d$ is $f=r d^2 = k d^2/n.$  Codes with a planar layout and strictly nearest neighbor connectivity are limited to $f\le c$, with $c$ a code dependent constant (\cite{Bravyi2010}). The surface code  saturates this bound with $d\sim \sqrt n$. One of the major topics in current QEC research is the development of more efficient codes that break the $f=c$ bound. This requires either higher dimensionality   which is difficult to implement ($d$ is upper bounded by $n/k$ in three dimensions), or relaxing the requirement of strictly local qubit connectivity. New families of quantum low density parity check (qLDPC) codes that utilize longer range, nonlocal qubit connections have been developed in recent years. These qLDPC codes are of particular interest for neutral atom implementations since long range connectivity is possible either through physical motion of qubits in optical tweezers (\cite{Beugnon2007,Bluvstein2024}) or in a static architecture using long range Rydberg interactions (\cite{Poole2025a, Pecorari2025}). 

 The availability of long range interactions has enabled alternative  implementations of entangling gates at the logical level. For the surface code, or any Calderbank-Shor-Steane (CSS) code, logical {\sf CNOT} gates can be performed by lattice surgery (\cite{Horsman2012}) which requires $d$ rounds of measurements for a distance $d$ code, or transversally (\cite{Shor2026}). Since neutral atom measurement times are orders of magnitude slower than gate times, and long range connectivity is available, transversal implementation of logical gates is a natural choice. Additionally, the development of correlated and algorithmic decoding techniques for fault tolerance with transversal operations promises to decrease error rates and increase logical cycle rates (\cite{Cain2024,Bluvstein2024,HZhou2025}).

 Another recent development for QEC with neutral atom qubits is the introduction of erasure conversion techniques. Erasure errors refer to errors that occur at known locations, which are generally leakage errors that take the atomic population out of the qubit's computational subspace, or physical loss of the atom. Both types of error are common in neutral atoms and handling them correctly is crucial for reliable computation.  With the assumption of perfect ancilla measurements and correct decoding a code with distance $d$ can detect and correct $t=\lfloor(d-1)/2\rfloor$ errors (\cite{Calderbank1996}).  Leakage errors into states outside the computational basis can be pumped back into the computational basis and thereby converted to Pauli errors. An example of an approach that does this for alkali atoms without disturbing valid qubit states can be found in \cite{Carr2013}. However, much  better performance can be obtained by converting leakage errors into erasure errors at known locations since a distance $d$ code can correct $t_{\rm max}=d-1$ erasures (\cite{Grassl1997}). If physical errors are predominantly erasures, or can be converted into erasures, the code threshold can be significantly boosted, thereby giving an exponential improvement in logical performance.  The applicability of erasure conversion to AEL qubits was pointed out in \cite{YWu2022} and erasure conversion was subsequently demonstrated with $^{87}$Sr (\cite{Scholl2023}) and $^{171}$Yb (\cite{SMa2023}) atoms. The benefit of erasure conversion in circuits using logical qubits was shown in \cite{BZhang2026}. Detection of leakage or loss and conversion into erasure errors is also possible with alkali atoms (\cite{ICong2022,MChow2024,CCYu2026}).

\section{Experimental architectures}

Combining the capabilities described in the preceding sections enables quantum processing satisfying the full set of DiVincenzo criteria (\cite{DiVincenzo2000}).   The first demonstrations of multi-qubit algorithms with neutral atoms  were performed in 2022 (\cite{Graham2022,Bluvstein2022}). These demonstrations 
utilized quite different control paradigms. In (\cite{Graham2022}) qubits were arranged on a planar grid and site-specific gate operations were performed by scanning focused laser beams to desired sites. This approach was used to demonstrate preparation of multi-qubit entangled states, as well as  quantum phase estimation and quantum approximate optimization algorithms. In (\cite{Bluvstein2022}) qubits were also arranged on a planar grid, with the addition of the ability to transport atoms between different regions using movable tweezers. Doing so enabled connectivity beyond the range afforded by direct Rydberg interactions.

Different architectural and hardware solutions may be preferred depending on the algorithm and application. An important example is the digital quantum computation of the dynamics of systems involving many fermions, such as arise in quantum chemistry and materials problems. Such calculations incur a large overhead in representing spins with qubits. The overhead can be reduced by using fermionic atoms as qubits (\cite{GonzalezCuadra2023}). 

A primary challenge in designing quantum computers is the need for logical operations between qubits that are not physically adjacent. This can be achieved using chains of swap operations to move quantum states across the processor. There is an associated cost in the errors accumulated in long chains and the time required. An alternative is to physically move atoms while preserving the quantum information encoded in internal states. This approach has been  demonstrated in several papers with Rb atoms based on different zones for memory, entanglement, and measurement (\cite{Bluvstein2024,Bluvstein2026}), and also adopted for demonstrations with logical qubits utilizing AEL atoms (\cite{Muniz2025b,BZhang2026}).  An alternative architecture relying on  gate operations with focused control beams, seeks to perform syndrome measurements in-place without motion (\cite{Miles2026}), while using nonlocal Rydberg interactions for moderate range logical operations, with motion used sparingly for only the longest range connections (\cite{Rines2025,Bergonzoni2026}). Less developed, yet novel  approaches include throwing and catching atoms to enable longer range connections (\cite{Hwang2023}) and three dimensional qubit arrays that provide greater connectivity, at the cost of more complex control (\cite{YWang2016,Kusano2025}).

Neutral atom systems have the potential to scale to many  qubits by leveraging the ability of optics to create large trap arrays in a straightforward manner. Nevertheless there are challenges in working at scale with atomic qubits. Imperfect vacuum conditions and the use of untrapped Rydberg states for entangling operations lead to atom loss. Atom motion for long range connectivity or transfer between different functional zones causes heating and eventual loss of atoms after many moves (\cite{Manetsch2025}). Repeated logical operations on data qubits lead to heating from photon recoils arising from optical control pulses, even without additional motion. The current state of the art is at the level of a few hundred operations, whereas beyond classical computations will require orders of magnitude longer circuits. Correction of atom loss and the ability to recool atoms midcircuit without stopping a quantum computation is therefore essential. 

Correction of atom loss is being addressed by developing continuously reloaded arrays that separate the production of cold atoms from the qubit array. Loss detection is incorporated into error correction protocols and replacement atoms are brought in with optical tweezers as needed from a reservoir of precooled atoms. There have been several recent demonstrations of this type of system (\cite{Dinardo2016,Norcia2024,Gyger2024,YLi2025,Chiu2025}) and with further engineering it appears possible to scale to many thousands of qubits.  

Mitigation of heating, which if left unchecked will lead to reduced operational fidelity in deep circuits, is also necessary. One approach is to periodically swap the quantum data between different groups of qubits, combined with recooling of the source group. This type of sequence can be configured to implement dynamic codes (\cite{McEwen2023}). The required periodic cooling of subsets of the atoms, without decohering data stored in the other atoms,  can be performed in a natural way using two atomic species, as has been shown with Rb and Cs arrays (\cite{Singh2022,Miles2026}). Alternatively teleportation can be used as an integral part of the syndrome extraction and error correction cycle to map data qubits onto newly cooled and initialized replacement atoms(\cite{MChow2024,Bluvstein2026}).

\section{Outlook}

This contribution has been completed in 2026 at a time of very rapid developments in the field of quantum computing seen broadly, and in the neutral atom approach specifically. Progress is being made in several areas: larger atom arrays, faster measurements, and higher fidelity gate operations. These capabilities are being synthesized into specific and detailed architectures for fault-tolerant quantum computation. While the surface code has been a favored choice for  quantum error correction for many years (\cite{Fowler2012}), quantum low density parity check (qLDPC) codes that are much more efficient have grown in popularity (\cite{Breuckmann2021}). The increase in efficiency enabled by these codes comes at the cost of requiring nonlocality for error syndrome measurement circuits. 
This requirement is challenging to meet in systems where the interactions are limited to nearest neighbors on a planar grid. Neutral atoms, on the other hand, can provide connectivity either through long-range Rydberg gates (\cite{Bergonzoni2026}), or via transport of blocks of atoms (\cite{Bluvstein2022,Bluvstein2026}).

A useful digital quantum computer will require  hardware aware compilation tools that minimize space and time resource requirements.
The availability of long range connectivity in neutral atom arrays opens up new possibilities for synthesizing optimized architectures for solving hard computational problems (\cite{Cain2026,SKhan2026}). A complete and detailed design for a large scale  optimized architecture has not yet emerged, and is an area of active research.
As minimization  of non-Clifford gate counts is itself an NP  hard computational problem (\cite{vandeWetering2023}), progress has focused on heuristic  solvers for neutral atom circuit compilation (\cite{Baker2021,DBTan2024,Schmid2024}).

Despite the impressive progress in scaling arrays of individually controllable qubits (see Fig. \ref{fig.arrays}), there are steep engineering challenges in scaling a single atom array to more than $10^5$ qubits. Large optical trap arrays are typically generated with spatial light modulators that have a finite number of controllable pixels. Scaling to arrays of $10^5$ sites or more is beyond the capability of available modulators. Custom optical components, including metasurfaces, are being developed to meet this challenge (\cite{Holman2026,CFang2026}).  A related challenge is thermal management of the requisite optical power, that necessarily increases with the array size. Even at the level of $\sim10,000$ site tweezer arrays, thermal management is a noticeable issue (\cite{Manetsch2025}). Atomic qubit arrays benefit from the dynamic reconfigurability of optical control, as regards both qubit positioning and application of logical operations. Realization of control with high space-bandwidth product is limited by available optical devices (\cite{Romer2014}). Extending fast control to larger arrays will require either cumbersome multiplexed solutions that combine many individual optical modulators, or development of new devices and architectures (\cite{Menssen2023,Graham2023,BZhang2024}).

An alternative to scaling a single, large array may rely on  a modular architecture with multiple arrays linked by photonic channels. 
Bell pairs shared across arrays can be used to teleport quantum states and gate operations, thereby enabling distributed circuit execution.  
Such systems  are under development (\cite{Young2022,YLi2024,Sinclair2025,Sunami2025b}) but are still at an early stage,  and lag behind operations within a single array in terms of both fidelity and rate. 

Despite the challenges enumerated above, the progress in neutral atom quantum computing in recent years has been spectacular. Many physicists, engineers, and computer scientists are engaged in overcoming the challenges   and  we may expect continued rapid progress towards broadly useful quantum computing with neutral atoms.  

\newpage

%\cite{Baker2021,HWang2024,Stade2025}

%\section{Conclusion}

\bibliographystyle{elsarticle-harv} 
\bibliography{qc_refs,rydberg,atomic,optics,saffman_refs}

\end{document}